# Las relaciones científicas entre Uruguay y Argentina en la década de 1920 a 1930: las ciencias exactas

Alejandro Gangui (1, 2) y Eduardo L. Ortiz (3)

(1) Universidad de Buenos Aires, Facultad de Ciencias Exactas y Naturales, Argentina.
(2) CONICET - Universidad de Buenos Aires, Instituto de Astronomía y Física del Espacio (IAFE), Argentina.
(3) Imperial College, London, U.K.

**Resumen**

La carrera científica del físico teórico uruguayo Enrique Loedel Palumbo en la Argentina ilustra el intenso intercambio intelectual que existió entre estos países vecinos en el primer tercio del siglo XX. En este trabajo discutimos brevemente la formación de este científico en el Uruguay, su posterior incorporación al Instituto de Física de La Plata, en Buenos Aires, y sus primeros pasos en la investigación bajo la tutela del profesor alemán Richard Gans.

**Introducción**

Presentaremos en primer lugar el ambiente liberal de la ciudad de Montevideo a comienzos del siglo pasado, en el que Loedel Palumbo desarrolló sus primeros estudios. Fue como alumno del Curso Preparatorio para el ingreso a la Universidad de la República que Loedel Palumbo entró en contacto con un personaje clave para su historia posterior, el Ing. Octavio Hansen, quien en esos años era el encargado de dictar los cursos de física en esa universidad. Hansen alentó al joven estudiante a avanzar en sus estudios de física y a publicar su primera monografía, en la que da cuenta de sus investigaciones personales en el campo de esa ciencia. Este último es un documento de considerable interés para el estudio de la personalidad científica de Loedel Palumbo, quien entonces contaba con tan sólo 19 años de edad.

En una segunda etapa, el estudiante se trasladó a la Universidad de La Plata, cuyo Instituto de Física había sido generosamente dotado. En La Plata estudió, simultáneamente, el doctorado en física en la Facultad de Ciencias y el profesorado en ciencias en la Facultad de Humanidades. Más tarde, esa doble experiencia le permitiría, a Loedel Palumbo, contribuir substancialmente a la actualización y a la mejora del nivel de los textos de enseñanza de las ciencias exactas dirigidos a la juventud. A su vez, esos aportes ayudarían al autor, al igual que a otros intelectuales de esos años, a sobrevivir económicamente durante diversos períodos en los que fue separado del medio universitario debido a interferencias políticas a nivel nacional.

En el Instituto de La Plata, Loedel Palumbo fue discípulo del reconocido físico alemán Richard Gans, de quién recibió una rigorosa formación como investigador científico. En su tesis de doctorado, que concluyó en 1925, se ocupó de una posible visualización de una molécula orgánica compleja en términos de sus cargas eléctricas. Los resultados de esos estudios fueron luego publicados por la prestigiosa revista alemana *Annalen der Physik*, en 1926.

En marzo-abril de 1925, cuando Albert Einstein visitó la Argentina, Loedel Palumbo no había finalizando aún su doctorado. Sin embargo, la Academia de Ciencias de Buenos Aires lo invitó a contribuir, con preguntas específicas, a una sesión especial que esa institución ofrecería al ilustre visitante (Gangui y Ortiz, 2014). Como consecuencia de sus intercambios con Einstein Loedel Palumbo completó un segundo trabajo -ahora sobre temas de relatividad general- que, nuevamente, publicó en una revista alemana, esta vez en *Physikalische Zeitschrift* (Loedel Palumbo, 1926c).

En la segunda mitad de la década de 1920 Loedel Palumbo estaba integrado, tanto al cuerpo docente de la Universidad de La Plata como a diversos institutos de enseñanza secundaria de esa ciudad. Su posterior trayectoria docente nos permite hacer algunas reflexiones acerca del sistema de acumulación de cargos que era prevalente en esos años en la Argentina. Este es un tema que aún estamos investigando (Gangui y Ortiz, 2018), pero lo que es seguro es que, para lograr reunir un salario adecuado este "sistema" obligaba a los investigadores a acumular posiciones de enseñanza en diferentes institutos, a veces situados en ciudades distantes; hicieron falta muchos esfuerzos -y tiempo- para que se arraigara el régimen de dedicación exclusiva que conocemos hoy.

**Los años en Montevideo**

Enrique Loedel Palumbo nació en la capital uruguaya el 29 de junio de 1901 en el seno de una familia culta. Sus padres, Juan Edoardo Loedel (1869-1957) y Emilia Palumbo (1869-1962), habían nacido ambos en Montevideo. Un antepasado de su padre fue uno de los principales propulsores de la incorporación del sistema métrico decimal en Uruguay (Loedel, 1864). Su madre, que era maestra y fue luego directora de una escuela en los alrededores de Montevideo, tuvo considerable influencia en el desarrollo intelectual de sus dos hijos: Enrique y Emilia Zoraida.

Loedel Palumbo ha relatado (Loedel Palumbo, 1940, p. 1) que descubrió su vocación científica y su interés por la física muy tempranamente, a los 14 años. También ha hecho referencia al papel que jugó su madre en el desarrollo de esos intereses intelectuales. A partir de aquella edad comenzó a formar en su casa un pequeño laboratorio en el que realizó experiencias científicas, algunas de ellas interesantes, registrando cuidadosamente sus resultados (Xenus, 1920).

Luego de terminar su bachillerato de cuatro años cursó los dos años de "estudios preparatorios" de ingeniería en la Universidad de la República, en Montevideo, donde conoció al Ing. Hansen (1876-1926), quien entonces era profesor de física en la Facultad de Matemáticas. Hansen es una figura interesante, de la que nos ocuparemos a continuación. Por su parte, la hermana de Enrique, Emilia Zoraida Loedel Palumbo (1896-1933) mostró también, desde su infancia, un singular talento para las matemáticas; más tarde, fue la primera mujer uruguaya graduada en ingeniería. Como profesional Emilia desarrolló una intensa actividad en el Ministerio de Obras Públicas de su país y, en mérito a sus contribuciones, una calle de Montevideo lleva hoy su nombre (Loedel Palumbo, 1927; Social Progress, 1928).

**Hansen, maestro de Loedel Palumbo**

El Ing. Hansen, cuyo padre también fue ingeniero, y cónsul de Dinamarca en la capital del Uruguay, es un personaje representativo del mundo cultural y científico de Montevideo en las primeras décadas del siglo XX, que fueron los años de infancia y juventud de Loedel Palumbo. Muy brevemente ahora mencionaremos algunos aspectos de la personalidad de Hansen (Mussio, 1926), y del medio en el que él y sus colegas desarrollaron sus actividades docentes en ese período.

En el último tercio del siglo XIX, el Ateneo del Uruguay, fundado en 1877, continuaba la labor educativa y formativa que había iniciado en Montevideo el Club Universitario. Este último había ayudado a crear un colegio, que luego fue sostenido por el Ateneo conjuntamente con un grupo de sociedades culturales de esa ciudad. Entre las sociedades que cooperaban con el Ateneo se contaba también una Sociedad Científica, contemporánea de su homónima de Buenos Aires (la Sociedad

Científica Argentina, fundada en 1872), y una Sociedad Amigos de la Educación Popular, fundada algo antes, en 1868.

En esos años, este grupo de instituciones brindó apoyo al desarrollo de la educación popular, elemental y superior, y también auspició la introducción de capítulos contemporáneos de la ciencia en la enseñanza. El Ateneo apoyó las corrientes del liberalismo racionalista, que llegaron al Uruguay hacia el último tercio del siglo XIX; en un momento de intenso debate cultural, su tendencia fue claramente laicista. Como se ha señalado (Zum Felde, 1987, I: 170-81), con sus trabajos y esfuerzos, el Ateneo y sus sociedades hermanas contribuyeron a la modernización del pensamiento intelectual del Uruguay, particularmente a partir de la década de 1880.

El Ateneo de Madrid, fundado en 1835 (Villacorta Baños, 1978) fue, sin duda, un modelo cuya influencia sobre el Ateneo del Uruguay, y sobre otras instituciones del Río de la Plata, es innegable. Aquél se autodefinía como una "Sociedad científica, literaria y artística" que se proponía "difundir las ciencias, las letras y las artes por todos los medios adecuados, y favorecer, dentro de su seno, el desarrollo de Agrupaciones que se propongan realizar la investigación científica y el cultivo de las artes y de las letras". Como ocurriría luego con las sociedades rioplatenses, sus actividades culturales estaban abiertas al público educado de Madrid.

Por otra parte, en 1869 el joven educador José Pedro Varela (1846-1879) fundó en Montevideo una escuela liberal y laica a la que dio el nombre de otro destacado maestro uruguayo, Elbio Fernández (1842-1869), fallecido en plena juventud. Ambos educadores, como antes Domingo F. Sarmiento (1811-1888) en la Argentina, admiraban los avances que la educación popular había alcanzado en los Estados Unidos, particularmente a través de las ideas y de la obra de Horace Mann (1796-1859), y consideraban a Benjamin Franklin (1706-1790) como uno de los principales inspiradores de esa corriente.

Sin duda alguna, esos educadores rioplatenses se inspiraban también en las ideas propagadas desde Madrid por la Institución Libre de Enseñanza (Jiménez Landi, 1973), fundada por Francisco Giner de los Ríos (1839-1915), juntamente con un grupo destacado de intelectuales liberales, en 1876. Su influyente *Boletín* facilitó una difusión amplia de sus ideas, tanto en España como en la América Hispana, donde fue leído con interés.

Hacia 1880, entonces, la Escuela Elbio Fernández contaba con el auspicio del Ateneo uruguayo y recibía, también, apoyo material de la Sociedad Amigos de la Educación Popular. En esos años había alcanzado ya considerable renombre por la calidad de la enseñanza que impartía y, en consecuencia, por el éxito de sus ex-alumnos. Varias personalidades jóvenes destacadas de la cultura uruguaya, por ejemplo el eminente escritor modernista -y político por el Partido Colorado- José Enrique Rodó (1871-1917), habían pasado por las aulas de esa escuela.

Octavio Hansen, que fue también uno de sus alumnos, inició sus estudios en un período de auge intelectual del Ateneo y, una vez finalizados, ingresó en la muy recientemente creada Facultad de Matemáticas y Ramas Anexas, dependiente de la universidad local, donde estudió la nueva carrera de ingeniería. Una vez graduado adquirió una vasta experiencia profesional: fue el primer director de Obras Municipales y a él se debe el diseño de la rambla de Montevideo, una de las principales atracciones arquitectónicas de esa ciudad y hoy en la lista "tentativa" de patrimonio mundial de la UNESCO. Además, formó parte del grupo que promovió la creación de la Asociación de Ingenieros del Uruguay, de la que fue elegido su tercer presidente, en 1909-10.

Hansen es parte de la primera generación de profesionales universitarios uruguayos que, en números ponderables, comenzaron a hacer entrar a su país en la escena de la técnica contemporánea. Esos jóvenes graduados se esforzaron por actualizar las estructuras tecnológicas del Uruguay intentando, de este modo, reducir la dependencia de los servicios de especialistas extranjeros. Algunos de ellos se asomaron también al mundo de la ciencia pura.

En 1905, sin abandonar su carrera profesional en la ingeniería urbana, Hansen fue designado profesor de física y, entre 1909 y 1913, actuó como miembro del Consejo Directivo de la Universidad de la República. Justamente en esos años se discutía la reorganización de la universidad y, también allí, Hansen jugó un papel destacado en los debates promoviendo la modernización de los estudios científicos en la Facultad de Ingeniería. Hasta ese momento la física no había logrado aún ganar un lugar propio dentro de los estudios de la carrera de ingeniería; sólo ocupaba una parte muy reducida en el plan de estudios básicos, en el ya mencionado "curso preparatorio". En esos años, en debates sobre la educación a nivel universitario, Hansen sostuvo que la enseñanza de la física debía "ser ampliada, especialmente en el calor, la electricidad y la meteorología" aconsejando agregar "la acústica para los estudiantes de Arquitectura y la óptica para los de Agrimensura" (Hansen, 1912, p. 41).

A grandes rasgos, este era el ambiente intelectual que el joven estudiante Loedel Palumbo encontró en Montevideo al ingresar a su universidad. Una vez en el preparatorio estableció un contacto estrecho con Hansen, quien lo alentó a continuar con sus estudios y, como vimos, también a publicar los resultados de sus indagaciones personales: así surgió el primer libro que mencionamos antes (Loedel Palumbo, 1920), en el que describió los resultados de sus primeros experimentos.

En esa obra el joven autor dio forma a sus ideas sobre varios tópicos de la física, ocupándose de las características geométricas de la vena líquida y del estudio de la teoría de los espejos esféricos; para este último tema utilizó ideas básicas de la geometría proyectiva que había aprendido -poco tiempo antes- en el curso preparatorio. Propuso también una técnica para la determinación de la velocidad de la luz, describió un fotómetro y un telégrafo de su invención y, finalmente, discutió la imposibilidad del movimiento continuo.

Por otra parte, Hansen propuso que se ofreciera a los alumnos del preparatorio asistir a cursos alternativos a los oficiales, con el objeto de elevar y modernizar el nivel de la enseñanza. Con esa idea, sugirió que Loedel Palumbo, a pesar de su extremada juventud y de ser aún un estudiante, fuera encargado de dictar un "curso libre" de física en el preparatorio, paralelo al curso oficial, pero más moderno y avanzado.

**Loedel Palumbo en La Plata**

Al finalizar el curso preparatorio, posiblemente por consejo de Hansen, Loedel Palumbo dejó Montevideo y se registró como alumno de la Licenciatura en Ciencias Físico-Matemáticas en la Facultad de Ciencias Físico-Matemáticas de la Universidad Nacional de La Plata (UNLP) (Loedel Palumbo, 1940, p. 2).

En ese momento esa universidad tenía el laboratorio de física más modernamente dotado de la región. Pero años antes, y poco después de que se fundara una universidad, aún provincial, en la nueva ciudad de La Plata, el ingeniero uruguayo Teobaldo Ricaldoni (1873-1923) había sido designado su primer profesor de física y se le había encargado equipar su laboratorio. Para ello adquirió una colección amplia de instrumentos de física, principalmente de demostración, en una conocida fábrica de instrumentos científicos de Alemania.

En 1909, en una nueva etapa iniciada por la nacionalización de esa institución, la nueva Universidad Nacional de La Plata creó un Instituto de Física orientado específicamente hacia temas de la física moderna. El antiguo laboratorio reunido por Ricaldoni fue puesto bajo la dirección del físico alemán Emil Bose (1874-1911), que había sido contratado especialmente en Alemania. Bose lo instaló en un edificio nuevo y lo reorientó hacia la enseñanza avanzada y la investigación original en física. Para alcanzar esos objetivos Bose creó una Licenciatura en Física e inició el dictado de cursos regulares de física a un nivel moderno y avanzado. Esos cursos produjeron los primeros graduados en física de la Argentina, y fueron ellos quienes, en gran medida, dominaron el panorama de esa ciencia por un período prolongado, que se extiende hasta mediados del siglo XX.

Inicialmente Ricaldoni continuó a cargo de los cursos básicos de física. Sin embargo Bose deseaba elevar el nivel de los estudios, e hizo intentos para modificar la situación. Lamentablemente, Bose falleció en La Plata en 1911, muy poco después de completar la instalación del nuevo laboratorio de física. Su sucesor fue otro alemán destacado, Richard Gans (1880-1954), quien dirigió el Instituto de Física hasta 1925. Gans era un físico de renombre que había contribuido al desarrollo de las teorías modernas del magnetismo; en La Plata contribuyó a formar un grupo amplio de físicos argentinos jóvenes y, a la vez, retuvo a aquellos cuyo entrenamiento había sido iniciado por Bose.

Durante la dirección de Gans varios alumnos comenzaron a publicar los resultados de sus investigaciones en una revista oficial de la universidad, fundada por Bose y refundada por Gans en 1913, llamada *Contribución al Estudio de las Ciencias Físicas y Matemáticas*.[1] Algunos de esos estudiantes comenzaron a colaborar también en revistas alemanas de prestigio.

Ese fue el singular ambiente de renovación cultural que Loedel Palumbo encontró en La Plata, una vez que cruzó el ancho río desde su Montevideo natal. Simultáneamente con su inscripción en la Licenciatura en Física, en la Facultad de Ciencias Físico-Matemáticas, se registró como estudiante del Profesorado en Matemáticas y Física, que se cursaba en un ámbito diferente, la Facultad de Humanidades de la misma universidad. Es posible que este interés por la educación, que como sabemos Loedel Palumbo conservó toda su vida, haya sido alentado primeramente por su madre y luego por su maestro Hansen.

En los años en los que Loedel Palumbo inició sus estudios en la Facultad de Ciencias los cursos fundamentales de física y matemática estaban a cargo de dos profesores contratados: el antes nombrado Gans y el matemático italiano Ugo Broggi (1880-1965), quien se había doctorado en Göttingen en 1907 (Broggi, 1907) trabajando en tópicos de probabilidad bajo la dirección de David Hilbert (1862-1943). Estos maestros tenían como asistentes a dos antiguos estudiantes de La Plata: Teófilo Isnardi (1890-1966) y Ramón Loyarte (1888-1944). Ambos habían sido iniciados en la investigación por Bose y luego continuaron trabajando bajo la dirección de Gans. Tanto Isnardi como Loyarte, lo mismo que un tercer estudiante del grupo inicial, José Collo (1897-1968), fueron enviados a Alemania para que perfeccionaran su formación.

Los dos cursos básicos de física general, que servían para revisar, reforzar y ampliar lo aprendido en la enseñanza secundaria, pasaron de Ricaldoni a Gans, que contaba con la ayuda de su asistente Loyarte. Este último dictó también, en 1923 y 1924, un curso de física especial (Anuario UNLP, 1923, p. 44; 1924, p. 41) en el que consideró la mecánica, la conducción del calor y la termodinámica, concluyendo con un panorama de la mecánica estadística y la teoría de los quanta.

Además de los cursos básicos, Gans tenía también a su cargo el curso de Trabajos de Investigación en Física (Anuario UNLP, 1923, p. 41; 1924, p. 41; 1925, p. 90), un curso central dentro de la carrera, que abría horizontes hacia la investigación original en esa ciencia. En esos años colaboraba también con el Instituto de La Plata el matemático alemán Paul Frank (1874-1938), que dictaba un curso de física-matemática. En sus cursos de 1923-24 (que fueron aquellos que tomó Loedel Palumbo) Frank desarrolló temas de electrodinámica clásica (Anuario UNLP, 1923, p. 36; 1924, p. 35).

---

[1] Llamada también *Contribuciones al Estudio de las Ciencias Físicas y Matemáticas* y, a veces, *Contribución al estudio de las ciencias físicomatemáticas*. La anotaremos como *Contribución* en lo que sigue.

Los cursos básicos de análisis matemático estaban a cargo de Broggi y de Emilio Rebuelto. Este último era un ingeniero de origen español con buena formación en la matemática clásica; por muchos años enseñó en Buenos Aires, colaborando a menudo con el matemático Julio Rey Pastor (1888-1962) en el dictado de cursos básicos para estudiantes de ingeniería. Broggi y Rebuelto dictaban, alternativamente, los dos cursos de análisis matemático siguiendo un texto especialmente redactado para la UNLP (Broggi, 1919-1927). En 1923 y 1924, Broggi fue también el responsable de un curso de matemáticas superiores (Anuario UNLP, 1923, p. 50; 1924, p. 46) en el que se ocupó, principalmente, de temas de análisis real.

De los documentos se puede ver que, hacia 1925-26, la UNLP intentó ampliar el círculo de intereses dentro de la Facultad de Ciencias, particularmente en dirección al desarrollo de la matemática superior. La incorporación de Rey Pastor a esa facultad se efectivizó precisamente en 1926, cuando se le encargó el dictado de un curso avanzado (Anuario UNLP, 1926, pp. 41-42), aunque la influencia intelectual de ese matemático, y su relación con la UNLP, es más antigua. Sin embargo, en esos años Loedel Palumbo ya había finalizado sus estudios.

**La "forma geométrica" de una molécula**

En diciembre de 1923 Loedel Palumbo se graduó como Profesor de Física en la Facultad de Humanidades, lo que le permitía dictar clases en escuelas de enseñanza media. Como sabemos, a juzgar por los documentos que hemos estado analizando (Gangui y Ortiz, 2018), esta posibilidad tendría una importancia considerable en su vida profesional, y también personal.

Muy poco más tarde obtuvo su primera posición en la enseñanza: una cátedra de física en el Colegio Nacional de La Plata, donde inició sus cursos en abril de 1924. En el mes de enero de ese año contrajo enlace con María Angélica Gorlero, de 19 años, a quién describiría como su novia desde los 14 años (Loedel Palumbo, 1940, p. 1). También en 1924, un año antes de la visita de Einstein a la Argentina, fue designado conservador en el Gabinete de Física de la Facultad de Ciencias Exactas, Físicas y Naturales de la Universidad de Buenos Aires (más conocida como Facultad de Ingeniería, lo que reflejaba la centralidad de los estudios de ingeniería en esa institución). Quizás ésta no fuera la posición ideal para un físico teórico, pero es muy posible que el nombre de su designación fuera, simplemente, un requerimiento presupuestario de esa facultad. Él permaneció en ese cargo hasta 1928 (Loedel Palumbo, 1940, p. 2).

A fines de 1925 Loedel Palumbo completó la licenciatura en física y, paralelamente con sus estudios del último año de esa carrera, preparó su tesis de doctorado en física bajo la dirección de Gans, en esos meses aún director del Instituto de Física y, sin duda, el físico más destacado de la Argentina de esos años. El 14 de diciembre de 1925 defendió, con éxito, su tesis titulada: "Determinación de las constantes ópticas de la sacarosa por la investigación de la refracción, de la despolarización de la luz de Tyndall y de la polarización circular en soluciones acuosas de tal sustancia" (así lo indica Westerkamp, 1975, p. 97), y poco más tarde le fue otorgado el diploma de doctor en física.

En su tesis de doctorado Loedel Palumbo utilizó diversas técnicas ópticas, magnéticas y eléctricas para medir parámetros físicos; entre otros, índices de refracción, de polarización circular y de despolarización de la luz de Tyndall. El tema de sus estudios fue la molécula de sacarosa, y su objetivo intentar atribuir un significado geométrico a las constantes que había medido: en pocas palabras, deseaba conjeturar acerca de "la forma geométrica de la molécula" de sacarosa. Más tarde, estas aptitudes geométricas del autor le servirían para investigar la "forma geométrica de la superficie espacio-tiempo de dos dimensiones en un espacio Euclídeo de tres", que fue justamente el tema de la pregunta que planteó a Einstein en Buenos Aires, y que despertó el interés del visitante (Ortiz, 1995; Gangui y Ortiz, 2005; 2008; 2014).

Volviendo al estudio de la molécula de sacarosa, éste era un tema en el que Gans estaba seriamente interesado y sobre el que había publicado diversos trabajos en esos años, principalmente en la prestigiosa *Annalen der Physik* (Gans, 1921), pero también en otras revistas científicas alemanas y europeas.

A principios de 1926 Loedel Palumbo presentó un resumen de su tesis para su publicación en *Contribución*, que apareció en ese mismo año (Loedel Palumbo, 1926a). El título de la versión publicada en esa revista es más específico que el de su tesis: "Las constantes ópticas de la molécula de sacarosa. Su 'forma geométrica'". En ese trabajo indicó que, siguiendo a Gans, a través del estudio de una molécula particular, y utilizando herramientas ópticas, pretendía inferir la "forma" geométrica de esa molécula. Concluyó su trabajo expresando que "debemos imaginar la molécula de sacarosa, en su comportamiento óptico, como algo así como un elipsoide torcido helicoidalmente, para explicar con ello la polarización circular" (Loedel Palumbo, 1926a, p. 78).

En su publicación Loedel Palumbo agradeció a su director, no solamente por guiar su trabajo sino también por haber "recibido siempre [de él] inapreciables consejos y múltiples atenciones". En esos años Gans venía haciendo un esfuerzo serio por educar a sus alumnos, no solamente a hacer investigaciones originales, sino también a concretar sus resultados y presentarlos en una forma publicable, tanto en el país como en la prensa científica internacional. Y ese fue el caso de la tesis de doctorado de Loedel Palumbo: una versión más reducida, con un apéndice teórico de Gans, fue publicada en *Annalen der Physik* (Loedel Palumbo, 1926b).

Pero en 1925 Gans dejó la UNLP para hacerse cargo de una cátedra en Königsberg, de modo que Loedel Palumbo fue su último discípulo en esa universidad. Como regresaba a Alemania antes de que su alumno pudiera presentar su tesis, ya que éste estaba aún cursando el último año de su carrera y debía, primeramente, rendir sus exámenes finales, Gans dejó documentado que esa tesis contaba con su total aprobación. Sin embargo, su propuesta encontró ciertos reparos formales dentro de la facultad, que el Consejo Universitario prontamente removió (Archivo UNLP, 1926). Estos documentos del archivo de la universidad, lo mismo que la correspondencia personal de Margrete Heiberg de Bose (1865-1952) indican que, luego de la partida de Gans, surgieron diferencias serias entre algunos miembros del Instituto de Física que antes habían sido sus alumnos. En aquella correspondencia es posible detectar esas diferencias aun inmediatamente después del fallecimiento de Bose.

Gans debe haber tenido sus motivos para abandonar la ciudad de La Plata, ya que en ese período las condiciones de vida en Alemania eran, todavía, sumamente duras debido a las secuelas de la Primera Guerra Mundial, incluso para un profesor universitario. Precisamente alrededor de esos años, en 1923-1924, se registró el pico histórico máximo de emigración en el sentido contrario: de Alemania hacia la Argentina (Rinke, 2005, p. 30). Años más tarde, en un homenaje tributado a Gans, en celebración de sus 70 años y de su regreso a la Argentina una vez terminada la Segunda Guerra Mundial, su ex-alumno Enrique Gaviola (1900-1989) hizo referencia a que "Circunstancias lamentables para la física argentina alejaron a Gans de La Plata entre 1925 y 1947. Sus viejos alumnos nos sentimos regocijados por su vuelta" (Gaviola, 1950).

**Conclusiones**

En este trabajo hemos repasado la trayectoria científica de Enrique Loedel Palumbo, quien inició sus estudios en Montevideo, en un ambiente cultural elevado, donde fue alumno de Octavio Hansen, uno de los primeros profesores de física de la Universidad de la República. Por su parte, Hansen había sido formado en la Escuela Elbio Fernández, dentro de las tradiciones liberales y laicas que entonces predominaban en los sectores intelectuales de esa ciudad.

En los inicios del siglo XX el Instituto de Física de la Universidad de La Plata se convirtió en un fuerte polo de atracción de jóvenes con talento, de ambas orilla del Plata, interesados en hacerse un futuro en las ciencias físicas. Formados por el profesor alemán Richard Gans, algunos de esos jóvenes alcanzaron cierto relieve en el campo de la física.

Entre ellos se destacó Loedel Palumbo, quien inició sus estudios en Uruguay y luego, muy joven, se trasladó a La Plata. En esa universidad, en el campo de la investigación, Loedel Palumbo llegó a ser, posiblemente, el discípulo más destacado entre los que Gans formó en la Argentina. Como hemos resaltado en trabajos anteriores, durante la visita de Einstein en 1925 Loedel Palumbo fue, sin duda, el interlocutor local más destacado.

Pero contemporáneamente a la visita del padre de la relatividad -o más precisamente a la partida de Gans- las condiciones en el Instituto de Física cambiaron, lo que paulatinamente fue alejando a Loedel Palumbo de las aulas y laboratorios de La Plata, y lo llevó a canalizar sus intereses científicos y docentes en otros lugares y temas. Estas son tan sólo algunas de las hebras sueltas de un ovillo complejo que aún debemos investigar.

**Bibliografía**

*Anuario*, Universidad Nacional de La Plata. 'Anuario para el Año 1923', No. 13, La Plata: Universidad Nacional de La Plata, No. 57, 1923.

*Anuario*, Universidad Nacional de La Plata. 'Anuario para el Año 1924', No. 14, La Plata: Universidad Nacional de La Plata, No. 60, 1924.

*Anuario*, Universidad Nacional de La Plata. 'Anuario para el Año 1925', No. 15, La Plata: Universidad Nacional de La Plata, No. 62, 1925.

*Anuario*, Universidad Nacional de La Plata. 'Anuario para el Año 1926', No. 16, La Plata: Universidad Nacional de La Plata, No. 70, 1925.

Archivo de la Universidad Nacional de La Plata. Expedientes: a) No. 1026, Decano Julio Castiñeiras a Dr. Teófilo Isnardi, 11.01.1926; b) No. 2867, Dr. Ramón G. Loyarte a Decano Julio Castiñeiras, 21.05.1926; c) No. 2172, Decano Julio Castiñeiras a Dr. Ramón G. Loyarte, 24.11.1926, 1926.

Broggi, Ugo. Die Axiome der Wahrscheinlichkeitsrechnung. Tesis, Univeristät Göttingen, 1907.

Broggi, Ugo. *Análisis Matemático*. La Plata: UNLP, Facultad de Ciencias Físicas, Matemáticas y Astronómicas, 1919-1927.

Gangui, Alejandro y Ortiz, Eduardo L. Marzo-abril 1925: Crónica de un mes agitado: Albert Einstein visita la Argentina. *Todo es Historia* **454**: 22-30, 2005.

Gangui, Alejandro y Ortiz, Eduardo L. Einstein's Unpublished Opening lecture for his Course on Relativity Theory in Argentina, 1925. *Science in Context* **21** (3): 435-450, 2008.

Gangui, Alejandro y Ortiz, Eduardo L. Einstein en la Argentina: el impacto científico de su visita. In: Bruno, P. (ed.). *Visitas culturales en la Argentina, 1898-1936*. Buenos Aires: Biblos, pp. 167-190, 2014.

Gangui, Alejandro y Ortiz, Eduardo L. Trabajo en preparación, 2018.

Gans, Richard. Asymmetrie von Gasmolekeln. Ein Beitrag zur Bestimmung der molekularen Form, *Annalen der Physik* **370** (10): 97-123, 1921.

Gaviola, Enrique. Homenaje a Ricardo Gans en su 70° aniversario: Introducción. *Revista de la Unión Matemática Argentina* **14** (3)**:** 100-108, 1950.

Hansen, Octavio. Preparatorios para los estudios de Ingeniería. *Anales de la Universidad* **17** (21): 39-48, 1912.

Jiménez Landi, Antonio. *La Institución Libre de Enseñanza. Los orígenes*. Madrid: Taurus, 1973.

Loedel, E. *Manual del sistema métrico de pesos y medidas, exposición completa, teórica y práctica*. Montevideo, 1864.

Loedel Palumbo, Emilia Z. Entrevista con la Ing. Emilia Z. Loedel Palumbo. *Mundo Uruguayo*, Sept. 15, 1927.

Loedel Palumbo, Enrique. *Nuevos conceptos y aplicaciones sobre algunos puntos de física*. Montevideo: Imprenta El Siglo Ilustrado, 1920 (130 pp).

Loedel Palumbo, Enrique. Las constantes óptica de la molécula de sacarosa. Su "forma geométrica". *Contribución al estudio de las ciencias físicas y matemáticas* **5**: 53-78, 1926a.

Loedel Palumbo, Enrique. Optische und elecktrische Konstanten des Rohrzuckers. *Annalen der Physik* **384**: 533-549, 1926b. [Con un Apéndice de Gans].

Loedel Palumbo, Enrique. Die Form der Raum-Zeit-Oberfläche eines Gravitationsfeldes, das von einer punkt-förmigen Masse herrürt. *Physikalische Zeitschrift*, **27**: 645-648, 1926c.

Loedel Palumbo, Enrique. Ficha Bio-Bibliográfica. La Plata, junio 25, 1940. La Plata: Colegio Secundario de Señoritas, Universidad Nacional de La Plata, 1940.

Mussio, Agustín H. Sepelio del Ing. Octavio Hansen. *Renacimiento* **3** (72), Nov. 1926.

Ortiz, Eduardo L. A convergence of interests: Einstein's visit to Argentina in 1925. *Ibero-Americanisches Archiv* (Berlin) **20**: 67-126, 1995.

Rinke, Stefan. German migration to Argentina (1918-1933). In: Thomas Adam (ed.). *Germany and the Americas*. Vol. I. Santa Bárbara: ABC-Clio, 2005, pp. 27-31.

Social Progress (Uruguay). *Bulletin of the Pan-America Union* **52** (116): 112, 1928.

Villacorta Baños, Francisco. *El Ateneo de Madrid, círculo de convivencia intelectual (1883-1913)*, Madrid: Consejo Superior de Investigaciones Científicas, 1978.

Westerkamp, José Federico. *Evolución de las Ciencias en la República Argentina, 1923-1972, La Física, Vol. II*. Buenos Aires: Sociedad Científica Argentina, 1975.

Xenus. Un libro interesante. *La Razón*, Montevideo, 24 de noviembre de 1920.

Zum Felde, Alberto. *Proceso Intelectual del Uruguay*, Vols. I-III. Montevideo: Ediciones del Nuevo Mundo, 1987.